\newcommand{\beq}{\begin{equation}}
\newcommand{\eeq}{\end{equation}}
\newcommand{\bea}{\begin{eqnarray}}
\newcommand{\eea}{\end{eqnarray}}

\newcommand{\gsim}{\lower.7ex\hbox{$\;\stackrel{\textstyle>}{\sim}\;$}}
\newcommand{\lsim}{\lower.7ex\hbox{$\;\stackrel{\textstyle<}{\sim}\;$}}

\newcommand{\mrm}{\mathrm}

\documentclass[aps,prev,twocolumn,preprintnumbers,floatfix,nofootinbib]{revtex4-1}
\pdfoutput=1
\usepackage{graphicx}
\usepackage{epstopdf}
\usepackage{mathrsfs}
\usepackage{amssymb}
\usepackage{verbatim}
\usepackage{color}
\usepackage{multirow}

\def\stacksymbols #1#2#3#4{\def\theguybelow{#2}
    \def\vp{\lower#3pt}
    \def\sp{\baselineskip0pt\lineskip#4pt}
    \mathrel{\mathpalette\intermediary#1}}

\def\intermediary#1#2{\vp\vbox{\sp
     \everycr={}\tabskip0pt
     \halign{$\mathsurround0pt#1\hfil##\hfil$\crcr#2\crcr
              \theguybelow\crcr}}}

\def\be{\begin{equation}}
\def\ee{\end{equation}}
\def\bea{\begin{eqnarray}}
\def\eea{\end{eqnarray}}

\def\sp{\;\;\;,\;\;\;}

\def\mrm{\mathrm}

\def\lsim{\raise0.3ex\hbox{$\;<$\kern-0.75em\raise-1.1ex\hbox{$\sim\;$}}}
\def\gsim{\raise0.3ex\hbox{$\;>$\kern-0.75em\raise-1.1ex\hbox{$\sim\;$}}}

\def\inbar{\,\vrule height1.5ex width.4pt depth0pt}

\def\IC{\relax\hbox{$\inbar\kern-.3em{\rm C}$}}
\def\IQ{\relax\hbox{$\inbar\kern-.3em{\rm Q}$}}
\def\IR{\relax{\rm I\kern-.18em R}}
 \font\cmss=cmss10 \font\cmsss=cmss10 at 7pt
\def\IZ{\relax\ifmmode\mathchoice
 {\hbox{\cmss Z\kern-.4em Z}}{\hbox{\cmss Z\kern-.4em Z}}
 {\lower.9pt\hbox{\cmsss Z\kern-.4em Z}}
 {\lower1.2pt\hbox{\cmsss Z\kern-.4em Z}}\else{\cmss Z\kern-.4em Z}\fi}

\def\comment#1{}
\def\to{\rightarrow}

\def\u1x{U(1)_X}
\newcommand{\nc}{\newcommand}
\nc{\LL}{L}
\nc{\vv}{\tilde{v}}
\nc{\ccdot}{\!\cdot\!}
\nc{\gsm}{G_{SM}}
\nc{\vfive}{\mathbf{5}\oplus\mathbf{\overline{5}}}
\nc{\vten}{\mathbf{10}\oplus\mathbf{\overline{10}}}
\nc{\zhol}{Z^{\rm hol}}
\nc{\xfb}{\,{\rm fb}}

\begin{document}

%
%

\preprint{CPHT-RR095.1214}
\preprint{LPT--Orsay 14-88}
\preprint{UMN--TH--3415/13}
\preprint{FTPI--MINN--14/44}

\vspace*{1mm}

\title{Monochromatic neutrinos generated by dark matter and the see-saw mechanism}

\author{Emilian Dudas$^{a}$}
\email{Emilian.Dudas@cpht.polytechnique.fr}
\author{Yann Mambrini$^{b}$}
\email{yann.mambrini@th.u-psud.fr}
\author{Keith A. Olive$^{c}$}
\email{olive@physics.umn.edu}

\vspace{0.1cm}
\affiliation{
${}^a$ CPhT, Ecole Polytechnique, 91128 Palaiseau Cedex, France }
\affiliation{
${}^b$ Laboratoire de Physique Th\'eorique 
Universit\'e Paris-Sud, F-91405 Orsay, France.
 }
 \affiliation{
${}^c$ 
 William I.~Fine Theoretical Physics Institute, 
       School of Physics and Astronomy,
            University of Minnesota, Minneapolis, MN 55455, USA
}

\begin{abstract} 

We study a minimal extension of the Standard Model where a scalar field is coupled to the
right handed neutrino responsible for the see-saw mechanism for neutrino masses.
In the absence of other couplings, below 8 TeV the scalar $A$ has a unique decay mode $A \rightarrow  \nu \nu$, $\nu$
being the physical observed light neutrino state. Above 8 (11) TeV, the 3-body (4-body) decay modes dominate. Imposing constraints on neutrino masses $m_\nu$ from atmospheric and solar
experiments 
implies a  long lifetime for $A$, much larger than the age of the Universe, making it a natural dark matter candidate.
Its lifetime can be as large as $10^{29}$ seconds, and its signature below 8 TeV would be a clear monochromatic neutrino signal, 
which can be observed by ANTARES or  IceCube.   Under certain conditions, the scalar $A$ may be viewed as a Goldstone
mode of a complex scalar field whose vacuum expectation value generates the Majorana mass for $\nu_R$. 
In this case, we expect the dark matter scalar to have a mass $\lesssim 10$ GeV.

\end{abstract}

\maketitle


\maketitle


\setcounter{equation}{0}



\section{Introduction}

The Standard Model (SM) of particle physics is now more than ever motivated by the 
recent discovery of the Higgs boson at both the ATLAS \cite{HiggsATLAS} and CMS
\cite{HiggsCMS} detectors. However, there are still two missing pieces in this elegant picture: 
the nature of dark matter (DM) and the origin of neutrino mass.
Despite the fact that a window for low mass dark matter candidates (below 100 TeV) seems favored 
by an upper bound coming for perturbative unitarity \cite{Griest:1989wd}, no evidence has been found after
many years of experimental searches \cite{DDexp}.
On the other hand, the presence of new physics at an intermediate scale seems motivated
by the stability of the Higgs potential \cite{EliasMiro:2011aa, OlegHiggs}, the see-saw  mechanism \cite{seesaw,seesaw2}, leptogenesis \cite{FY,leptogenesis} 
or reheating processes. Added to the fact that a super--heavy DM, or WIMPZILLA \cite{WIMPZILLA} could be produced by non--thermal processes
and explain the DM density in the universe, it seems natural to link the mechanism for generating a
neutrino mass with the dark matter question 
in a coherent framework at an intermediate scale.

An intermediate scale (of order $10^{10}$ GeV) will never be reached 
by an accelerator on earth. The 100 TeV collider is
still a state of mind project, whereas the ILC can reach, at most, a beyond the SM (BSM) 
scale of 100 TeV through precision measurement.
However, we know that energies as large as $10^{10}$ GeV are measured in ultra-high energy cosmic rays experiments
like the Auger observatory \cite{Auger}.  Recently, the IceCube collaboration claimed the detection of multi PeV ($10^{6}$ GeV) events \cite{Icecube},
giving the community some hope that an intermediate scale can be testable in the near future with these
types of experiments.

In this letter we show that a high energy neutrino signal can be associated with a long-lived scalar dark matter 
candidate. We show that this scalar can account for the dark matter in the Universe and 
moreover, its specific decay mode into two monochromatic neutrino states gives a clear signature detectable in present 
high energy detectors like IceCube \cite{Icecube}. 
We then attempt to relate this candidate with the pseudo-Goldstone mode of
a complex scalar field responsible for 
generating a Majorana mass in the right handed sector through dynamical symmetry breaking at an intermediate scale. 

The letter is organized as follow. After a description of the single scalar model  we analyze in section II, we compute its phenomenological
consequences and detection prospects in section III. In section IV, we relate this scalar as the Goldstone mode associated with generating the right-handed neutrino mass, necessary 
for the see-saw mechanism.  We draw our conclusions in section V.


\section{Dark matter and a standard see--saw mechanism}
\label{sec:model}

\subsection{The model}

\noindent
As a simple extension to the SM with a neutrino see-saw mechanism, we add a single real scalar field, $A$,
coupled to the right handed (sterile) sector. The Lagrangian can then be written as
\be
{\cal L} = {\cal L}_{\mrm{SM}} + {\cal L}_\nu + {\cal L}_A
\ee
with
\be
{\cal L}_\nu =  -(\frac{1}{2} M^R +\frac{i h}{\sqrt{2}} A) \bar \nu_R^c \nu_R
- \frac{y_{LR}}{\sqrt{2}} \bar \nu_L H \nu_R + h.c.
\label{Eq:lagrangian1}
\ee
and
\bea
&&
{\cal L}_A = - \frac{\mu_A^2}{2} A^2 - \frac{\lambda_A}{4} A^4  
\nonumber
\\
&&
-\frac{\lambda_{H A}}{4} A^2  H^2 + \frac{1}{2} \partial_\mu A \partial^\mu A
\label{Eq:lagrangian2}
\eea
where $H$ represents the real part of the SM Higgs field. 
Here, we have simply assumed that the right handed neutrino has a Majorana mass, $M^R$.
We will explore a dynamical version of this extension in section IV.

The scalar $A$ is massive and couples to the SM Higgs, but does not itself get a vacuum expectation value ($vev$). 
While there is no natural value for the mass scale $M^R$,   demanding gauge coupling unification in different schemes of SO(10) breaking
naturally leads to intermediate scales between $10^6-10^{14}$ GeV \cite{Fukugita:1993fr,moqz}. It seems then reasonable to expect that $M^R$ 
will lie in this energy range if one embeds our model in a framework where one imposes unification of the gauge couplings.
However, we will attempt to stay as general as possible\footnote{{We note that a similar framework has been used in \cite{Higgsstability} 
to stabilize the Higgs potential up to GUT scale.}}.  In the context of very light scalar $A$, of order a keV (though not considered in the present work),
some authors have looked at the effect of a decaying $A$ on the CMB \cite{valleold} and more recently the subleading effect of decays to photons 
 \cite{vallenew}.  

\subsection{The see--saw mechanism}

Once symmetry breaking is realized, the mass states in the neutrino sector are mixed in the current eigenstate basis. 
Diagonalization of the mass matrix leads to the well known see--saw mechanism. 
We can write the mass term
\bea
{\cal L}_\nu = -\frac{1}{2} \bar n~ {\cal M}~ n, ~~ \mrm{with} ~~ n = \left( \begin{array}{c} \nu_L  + \nu_L^c \\ \nu_R + \nu_R^c \end{array} \right)=\left( \begin{array}{c} n_1 \\ n_2 \end{array} \right) 
\nonumber
\eea
and 
\be
{\cal M} = \left(  \begin{array}{cc}  0 \ m_D \\ m_D \ M^R \end{array} \right),
\label{Eq:m}
\ee
with $m_D =y_{LR} v_H / \sqrt{2} $ ($v_H=246$ GeV being the Higgs $vev$). ${\cal M}$, being a complex symmetric matrix, can be diagonalized with the help of {\it one unitary matrix U}, 
 ${\cal M}=U m U^T$ with
 \be
 m = \left( \begin{array}{cc} m_1 & 0 \\ 0 & m_2 \end{array} \right) .
 \ee
From the diagonalization of {\cal M} 
\bea
&&
m_1= \frac{1}{2} \biggl[M^R - \sqrt{(M^R)^2 + 4 m_D^2 } \biggr] \simeq - \frac{m_D^2}{M^R} \simeq - \frac{y_{LR}^2 v_H^2}{2 M^R}
\nonumber
\\
&&
m_2=\frac{1}{2} \biggl[M^R + \sqrt{(M^R)^2 + 4 m_D^2} \biggr] \simeq M^R 
\label{Eq:seesaw}
\eea
 and the eigenvectors $N_1$ and $N_2$
 \be
 \left( \begin{array}{c} N_1 \\ N_2 \end{array} \right) \simeq \left(\begin{array}{c} n_1 - \theta ~ n_2 \\ n_2 + \theta ~  n_1 \end{array} \right)  
= \left( \begin{array}{c} \nu_L + \nu_L^c - \theta ~ (\nu_R + \nu_R^c) \\ \nu_R + \nu_R^c + \theta ~(\nu_L + \nu_L^c) \end{array} \right)
 \ee
where\footnote{Notice that $N_1$ and $N_2$ are Majorana like particles.} $\tan 2 \theta = - \frac{2 m_D}{M^R} $ implying $\theta \simeq \sin \theta \simeq - \frac{m_D}{M^R} = - \frac{y_{LR} v_H}{\sqrt{2} M^R}$.
Once the Lagrangian is expressed in terms of the physical mass eigenstates, one can compute the couplings generated by the symmetry breaking and their phenomenological consequences.  $m_1$ corresponds to the mass of the Standard Model neutrino. We will consider 
$m_1 \lesssim 1$ eV from cosmological constraints through the rest of the paper\footnote{We neglect the flavor structure of the SM neutrino
sector as it does not affect our main conclusions.}.

\section{Phenomenology}

\subsection{Generalities}

To study the consequences of the model, we first
rewrite the Lagrangian (\ref{Eq:lagrangian1}) in terms of the mass eigenstates, $N_{1,2}$, 
\bea
&&
 {\cal L}_\nu = -\frac{y_{LR}}{2 \sqrt{2}} H \left( \bar N_1 N_2 + \bar N_2 N_1 \right) - \frac{y_{LR} \theta}{\sqrt{2}} H\left(\bar N_2 N_2 - \bar N_1 N_1  \right)
 \nonumber
 \\
 &&
 - \frac{m_1}{2} \bar N_1 N_1 - \frac{m_2}{2} \bar N_2 N_2
 \label{Eq:lagrangian3}
 \\
 &&
 -i \frac{h}{\sqrt{2}} A \left( \bar N_2  \gamma^5 N_2 - \theta \bar N_1  \gamma^5 N_2 - \theta \bar N_2  \gamma^5 N_1  + \theta^2 \bar N_1 \gamma^5 N_1 \right)
 \nonumber
\eea

A look at the Lagrangian (\ref{Eq:lagrangian3}) implies some obvious phenomenological consequences of the coupling of the scalar to the neutrino sector.
First of all, the field $N_2$ is not stable through its decay $N_2 \rightarrow H N_1$ and cannot be the dark matter candidate as in the standard see-saw mechanism.  Secondly, the scalar $A$ is not stable, and
its dominant decay mode for $M_A\lesssim$ 8 TeV  is $A \rightarrow N_1 ~N_1$, as $M_{N_2} =m_2$ is of the order of $M^R$ and is for now assumed to be heavier than $A$. When we include $A$ as part of a dynamical mechanism for
generating the mass $M^R$, we will see that the mass of $A$ may be highly suppressed relative to $M^R$,
justifying a posteriori our assumption that $M_A < M_{N_1}$,
Moreover, because the coupling of $A$ to $N_1$ is of order of $h \theta^2$, $A$ is naturally long-lived,  and can be a good dark matter candidate as we will see below.
Its decay signature should be two ultra energetic monochromatic neutrinos which is a clear observable, and could be accessible to 
the present neutrino experiments like IceCube, ANTARES or SuperK as we will see below.
Above 8 TeV, the 3-body decay mode dominates and above 11 TeV, the 4-body decay mode dominates. 
In this case, the signature of decay is then no longer a monochromatic signal but a 
neutrino-spectrum as we will see in the next section.

\subsection{Neutrino flux}

\noindent
Before computing the flux of neutrinos expected on earth from the decay of the scalar $A$, let's first check if it can be a reliable
dark matter candidate, fulfilling $\tau_{A} > \tau_{\mrm{Universe}} = 10^{17}$ seconds\footnote{A recent study \cite{Audren:2014bca} showed
that taking into account the recent BICEP2 results, the real lifetime to consider should be $\gtrsim 10^{18}$ s.
However, the constraints from IceCube are much stronger ($\tau_A \gtrsim 10^{28}$ seconds for $M_A$ at the PeV scale) as we will see below.}. 
The 2-body decay width for $A \rightarrow N_1 N_1$ is given by
\be
\Gamma^2_A = \frac{10^{-38} h^2}{8 \pi}  \left( \frac{m_1}{\mrm{1~eV}} \right)^2 \left( \frac{10^{10}~\mrm{GeV}}{M^R}  \right)^2
M_A~
\nonumber
\ee
implying
\be
\tau_A \sim 1.6 \times 10^{12} h^{-2} \left( \frac{1 ~\mrm{eV}}{m_1} \right)^2 \left( \frac{M^R}{10^{10} ~\mrm{GeV}} \right)^2
\left( \frac{1~\mrm{TeV}}{M_A} \right)~~\mrm{[s]} ,
\label{Eq:taur21}
\ee
where we have taken for reference $m_1 \lesssim 1$ eV
as implied 
 from the solar and atmospheric constraints on neutrino masses. 
 As one can see, for a scalar mass of order 1 TeV, one can 
 obtain lifetimes in excess of the age of the Universe for $M^R \gtrsim 10^{13} h $ GeV,
making the scalar a potentially interesting dark matter candidate. 

However, it is important to check multi-body processes when $M_A  > v_H$. Indeed\footnote{The authors want to thank the referee for
having pointed out this possibility.} the 3-body process $A \rightarrow N_1 N_1 H $ or the 4-body decay $A \rightarrow N_1 N_1 H H$, through 
the exchange of a virtual $N_2$ becomes dominant. Under the approximation of massless final states (largely valid when $M_A \gg m_h$)
one obtains for the 3- and 4-body decay widths
\be
\Gamma^3_A = \frac{10^{-38} h^2}{3 \times 2^{8} \pi^3}  \left(\frac{m_1}{1~\mrm{eV}}  \right)^2 \left( \frac{10^{10}~\mrm{GeV}}{M^R} \right)^2
\left( \frac{M_A}{v_H} \right)^2 M_A
\ee
 and
\be
\Gamma^4_A = \frac{10^{-38} h^2}{9 \times 2^{14} \pi^5} \left( \frac{m_1}{1~\mrm{eV}} \right)^2 \left( \frac{10^{10}~\mrm{GeV}}{M^R} \right)^2
\left( \frac{M_A}{v_H} \right)^4 M_A
\ee 
which gives for the total width
 \bea
 &&
 \Gamma_A= \frac{10^{-38} h^2}{8 \pi} \left( \frac{m_1}{1~\mrm{eV}} \right)^2 \left( \frac{10^{10}~\mrm{GeV} }{M^R} \right)^2 M_A 
 \times
 \nonumber
 \\ 
 &&
 \left[ 1 + \frac{1}{96 \pi^2} \left( \frac{M_A}{v_H} \right)^2 + \frac{1}{18432 \pi^4} \left( \frac{M_A}{v_H} \right)^4\right]
 \eea
 From the expression above, we can easily compute that for $M_A \gtrsim 10 \pi v_H \simeq 8$ TeV, the three body, and then for 
 $M_A \gtrsim 11$ TeV the 4-body dominate the decay process. It will be interesting then to see what kind of constraints Icecube or ANTARES 
 can impose on the parameter space of the model\footnote{An earlier analysis in another context were proposed in
 \cite{PalomaresRuiz:2007ry}}.

The IceCube collaboration, recently released their latest analysis \cite{Icecube} concerning the (non)observation of ultra high
 energetic neutrino above 3 PeV. The PeV event rate expected at a neutrino telescope of fiducial volume $\eta_EV$ and nucleon number density
 $n_N$ from a decaying particle of mass $M_{DM}$, mass density
  and width $\Gamma_{DM}$ is \cite{Feldstein:2013kka}
 \bea
 &&
 \Gamma_{\mrm{events}}= \eta_E V \times n_N \times \sigma_N \times L_{\mrm{MW}} \times \frac{\rho_{DM}}{M_{DM}} \times \Gamma_{DM}
 \nonumber
 \\
 &&
 \simeq 3 \times 10^{59}~ \eta_E \frac{\Gamma_{DM}}{M_{DM}} ~\mrm{years^{-1}} ,
 \label{Eq:flux}
 \eea
where $\sigma_N$ ($\simeq 9 \times 10^{-34} \mrm{cm^2}$ at 1 PeV)
  is the neutrino--nucleon scattering cross section, $n_N$ is the nucleon number density in the ice
 ($n_N \simeq n_{\mrm{Ice}} \simeq 5 \times 10^{23}/ \mrm{cm^3}$), $L_{MW}$ is the rough linear dimension of our galaxy (10 kpc)
 and $\rho_{DM} \simeq 0.39 ~\mrm{GeV}~\mrm{cm^{-3}}$ 
 the milky way dark matter density (taken near the earth for the purpose of our estimate). The volume
 $V$ is set to be 1 $\mrm{km^3}$, which is roughly the size of the IceCube detector, whereas the efficiency coefficient 
 $\eta_E$ depends on the energy of the incoming neutrino and lies in the range $10^{-2}-0.4$ \cite{Veff}.

The neutrino--nucleon cross section is, however,  highly dependent on the scattering energy and
 the authors of \cite{Gandhi:1998ri} obtained
 \bea
 \sigma_N &&= 8 \times 10^{-36} \mrm{cm^2} \left( \frac{E_\nu}{\mrm{1 ~ GeV}} \right)^{0.363} 
 \nonumber
 \\
 &&
 = 6 \times 10^{-36}  \mrm{cm^2} \left( \frac{M_{DM}}{1 ~\mrm{GeV}} \right)^{0.363} .
 \label{Eq:sigman}
 \eea

 Implementing Eq.(\ref{Eq:sigman}) in the expression (\ref{Eq:flux}), and adding an astrophysical factor $f_{\mrm{astro}}\simeq 1$ corresponding
 to the uncertainty in the distribution of dark matter in the galactic halo, one can write
 \be
 \Gamma_{\mrm{events}} = 1.5  \times 10^{57}\eta_E f_{\mrm{astro}}  \frac{\Gamma_{DM}}{M_{DM}^{0.637}} ~~\mrm{years^{-1}} ,
 \label{Eq:gamma}
 \ee
 where $\Gamma_{DM}$ and $M_{DM}$ are expressed in GeV.
 Noticing that there is no background from cosmological neutrino at energies above 100 TeV, one can deduce the limit set by IceCube
 from the non-observation of events above 3 PeV. IceCube took data during 3 years, so asking $3 \times \Gamma_{\mrm{events}} \lesssim 1$
 one obtains for $f_{\mrm{astro}}=1$
  \be
 h^2 \left( \frac{M_A}{1~\mrm{GeV}} \right)^{4.363}  \eta_E \lesssim 3.7 \times 10^{-3} \left( \frac{1~\mrm{eV}}{m_1}\right)^2 \left( \frac{M^R}{10^{10}~\mrm{GeV}} \right)^2
 \ee
  If we take $M_A=1$ PeV, one obtains 
 $h \lesssim  8 \times 10^{-11}$  for $M^R \sim 10^{14}$ GeV, $\eta_E=0.4$ and $m_1=1$ eV.

\begin{figure}
    \begin{center}
    \includegraphics[width=3.in]{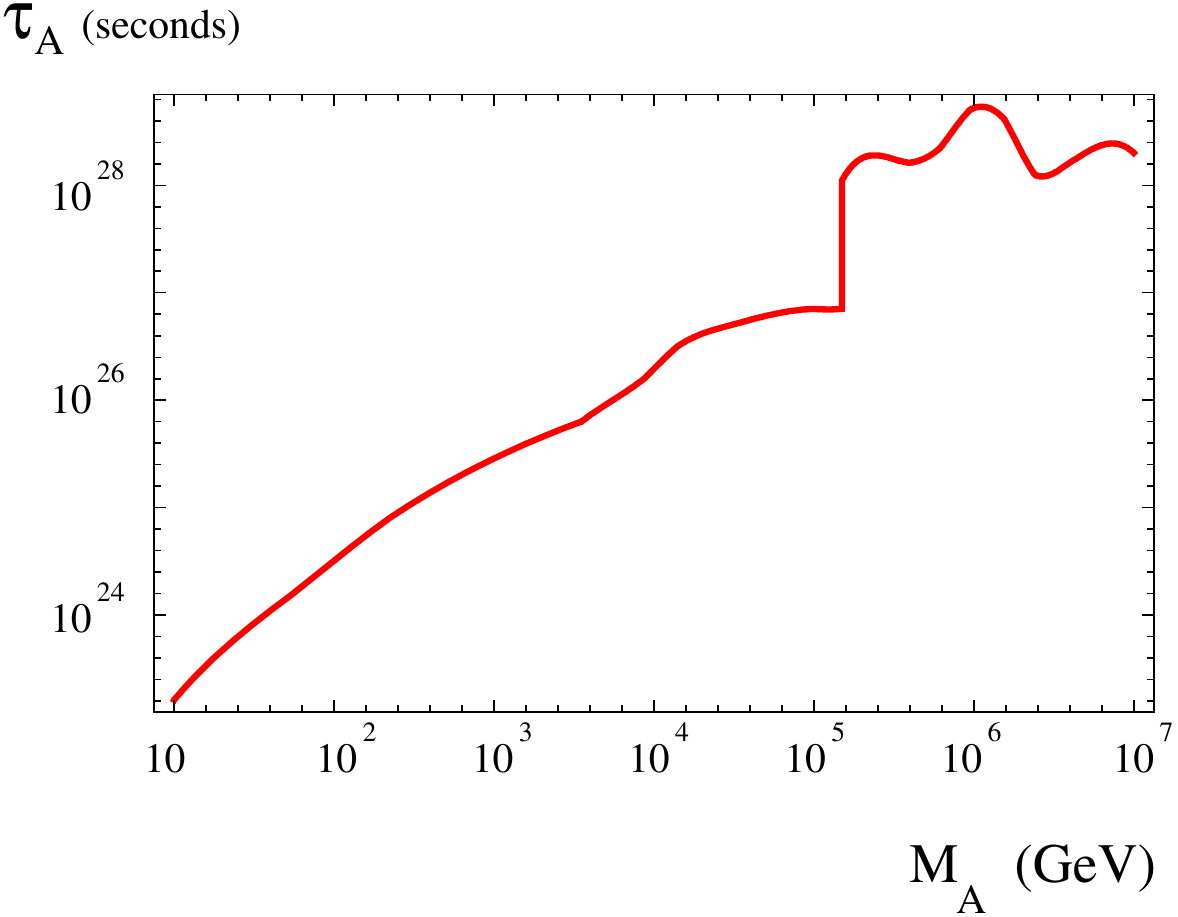}
              \caption{{\footnotesize
     Limit on the lifetime of $A$ in seconds as a function of its mass $M_A$ extracted from the combined constraints \cite{Covi:2009xn,Ibarra:2013cra,Rott:2014kfa} from SuperK \cite{superk},
ANTARES \cite{antares} and IceCube \cite{Abbasi:2011eq}}}.
\label{Fig:taulimit}
\end{center}
\end{figure}

\begin{figure}
    \begin{center}
    \includegraphics[width=3.in]{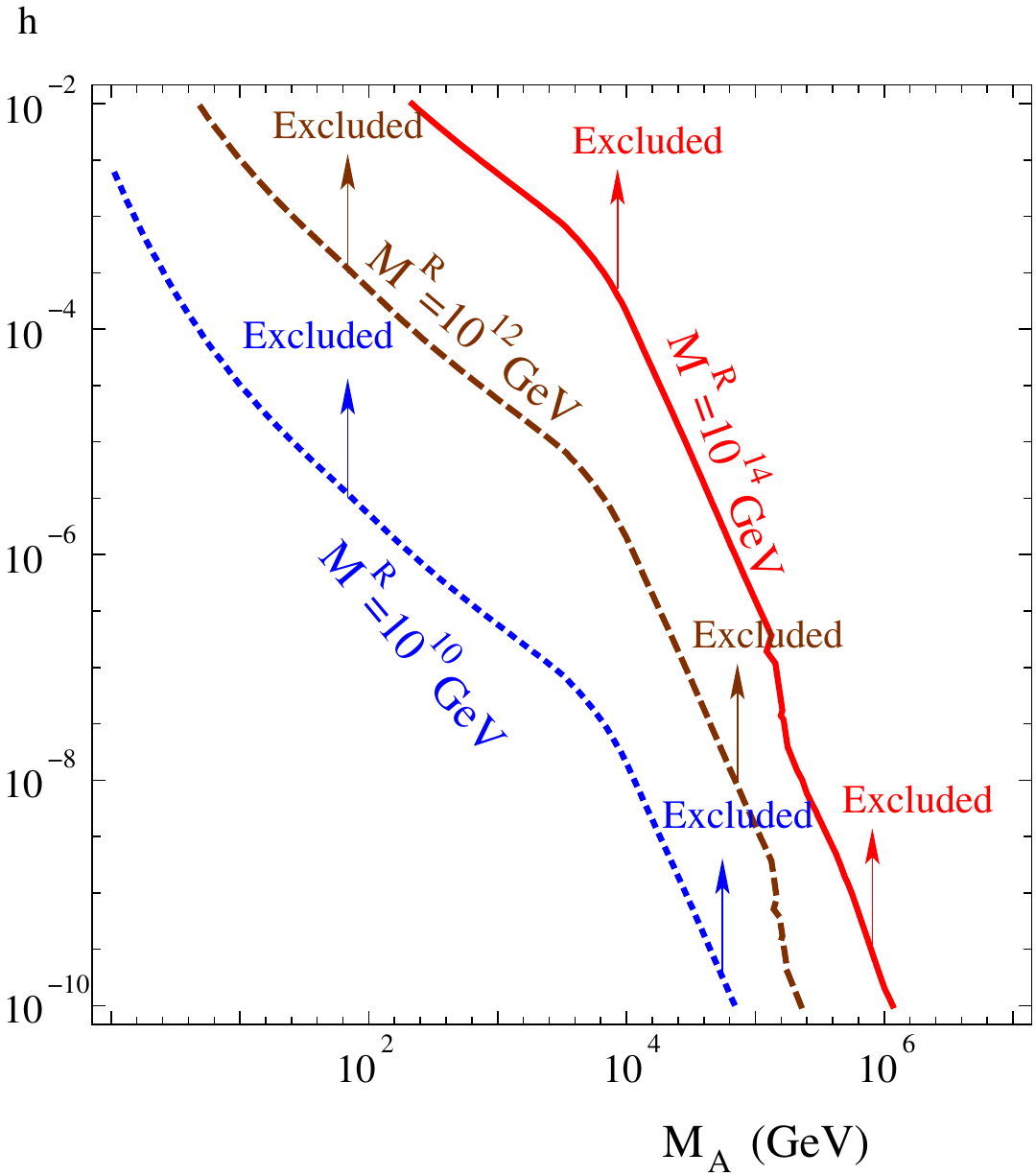}
              \caption{{\footnotesize
         Parameter space allowed in the plane ($M_A, h$) taking into account a combined analysis of IceCube and SuperK 
 for  different values of $M^R$ and $m_1=1$ eV. The regions above the lines are excluded.
}}
 \label{Fig:h}
\end{center}
\end{figure}

One can generalize our study to lower masses, down to the GeV scale, taking into account the combined constraints \cite{Covi:2009xn,Ibarra:2013cra,Rott:2014kfa} from SuperK \cite{superk},
ANTARES \cite{antares}  and IceCube \cite{Abbasi:2011eq}. The limit on the lifetime of $A$ as function
of $M_A$ is depicted in Fig. \ref{Fig:taulimit}.   
The resulting constraint in the  ($M_A, h$) parameter plane is shown in Fig. \ref{Fig:h} for different values of $M_R$. We see that natural values of $M^R$ ($\gtrsim 10^{12}$) GeV leads to upper limit on 
$h \lesssim 10^{-5}$, for $M_A > 1 $ TeV.

We note that despite the fact that a dark matter source for the PeV events of 
IceCube are less motivated since the discovery of the third event ``big bird", one can also 
compute the relation between $h$ and $M^R$ to observe the rate of 1 event per year for a 1 PeV dark matter candidate.
We obtain from eq.(\ref{Eq:gamma}) $h \simeq 1.3 \times 10^{-10}$ for $M^R=10^{14}$ GeV .

We also made a more detailed analysis, taking into account a simulated NFW galactic profile $\rho_{NFW}$ for the milky--way.
Our result differs from the constraint with $f_{astro}=1$ only by a factor of a few (2-3). Indeed, compared to an annihilating scenario, due to the lack of
quadratic enhancement in the signal, the role of the (better determined) local density is more prominent. We can thus anticipate 
little dependence of our conclusions on the specific galactic halo used for the analysis.
 

\section{Dark Matter and a Dynamical See-Saw}

\subsection{The model}

We would now like to ask whether or not, the scalar $A$ can be incorporated into
a dynamical mechanism for generating neutrino see-saw masses.
Instead of the coupling of $A$ to $\nu_R$ in Eq. (\ref{Eq:lagrangian1}), let us 
couple the right handed (sterile) sector to a complex scalar
field $\Phi = Se^{ia/v_S}$. We assume that $\Phi$ is responsible for the breaking of some global symmetry so that $S$ acquires a $vev$. 
Here,  we would like to stay as general as possible, 
and show that our framework can in fact be an illustration of any extension to the SM with dynamical breaking occurring at an intermediate scale. We now rewrite the Lagrangian as
\be
{\cal L} = {\cal L}_{\mrm{SM}} + {\cal L}_\nu + {\cal L}_\Phi
\ee
with
\be
{\cal L}_\nu =  - \frac{h}{\sqrt{2}} \bar \nu_R^c \Phi \nu_R
- \frac{y_{LR}}{\sqrt{2}} \bar \nu_L H \nu_R + h.c.
\label{Eq:lagrangian4}
\ee
and
\bea
&&
{\cal L}_{\Phi} = \frac{\mu_\phi^2}{2} |\Phi|^2 - \frac{\lambda_\Phi}{4} |\Phi|^4  
\nonumber
\\
&&
-\frac{\lambda_{H \Phi}}{4} |\Phi|^2 |H|^2 + \frac{1}{2} \partial_\mu \Phi \partial^\mu \Phi^*  .
\eea
The lagrangian above has a global $U(1)$ symmetry, under which the charges of
$\nu_R$, $\nu_L$ and $\Phi$ are $1$,$1$ and $-2$, respectively. 
After symmetry breaking generated by the fields $H$  and $\Phi$, one obtains in the heavy sector,
\be
\langle S \rangle = v_S = \frac{\mu_\Phi}{\sqrt{\lambda_{\Phi}}}~; ~~ S = v_S + s~; ~~M_S = \sqrt{2} \mu_\Phi,
\ee
and we denote by $A$, the argument of $\Phi$ after its magnitude is shifted by $v_S$. 
Note that our scalar field has been promoted to a Goldstone mode and is massless at tree level.
The right-handed mass, $M^R$ is now given by $h v_S/\sqrt{2}$. 


\subsection{Breaking to a discrete symmetry.}

If our U(1) symmetry was exact (prior to $\Phi$ picking up a $vev$), $M_A$ would remain
massless to all orders in perturbation theory.
In what follows, we will assume that the $U(1)$ symmetry is broken by nonperturbative effects down to a discrete $Z_N$ symmetry. It is actually standard in string theory that
all symmetries are gauged symmetries in the UV. Some of them nonlinearly realized, in the sense that under gauge transformations one axion 
$\tilde \theta$ has a nonlinear transformation
\be
A_\mu \to A_\mu + \partial_\mu \alpha \ , \ \Phi_i \to e^{i q_i \alpha} \Phi_i
\ , \ \tilde \theta \to \tilde \theta + \alpha \ , \label{zn1} 
\ee
and the lagrangian contains the Stueckelberg combination of a massive vector boson
\be
\frac{M^2}{2} (A_{\mu} -  \partial_\mu \tilde \theta)^2 \ . \label{zn2} 
\ee
Nonperturbative effects can generate operators of the form \cite{nonpert}
\be 
\frac{c_n}{M_P^{n-4}}e^{- 2 \pi N (t+i \frac{{\tilde \theta}}{2 \pi})} \prod_i \Phi_i \ , \label{zn3} 
\ee
where $t$ is a field which gets a vev and where $S_{\rm inst.} = 2 \pi N (t+i \frac{\tilde \theta}{2 \pi})$ can be interpreted as an instanton action. 
Nonperturbatively generated operators  (\ref{zn3}) are gauge invariant, provided that
$\sum_i q_i = N$.  The gauge field $A_\mu$ which eats the axion $\tilde \theta$ and the field $t$ can be very heavy and decouple at low energy. At low energies therefore one gets an effective operator
\be 
e^{- \langle S_{\rm inst.} \rangle } \frac{c_n}{M_P^{n-4}} \prod_{i=1}^n \Phi_i \ \ , \label{zn4} 
\ee
with $c$ a numerical coefficient. 
At low energy, even though the $U(1)$ gauge symmetry is broken, one obtains a remnant
$Z_N$ symmetry. Due to its gauge origin, it satisfies anomaly cancellation conditions \cite{Ibanez:1991hv}. If the original gauge symmetry was anomaly-free (which is realized if the axionic coupling to gauge 
gauge fields as ${\tilde \theta} F {\tilde F}$ is absent),  then anomalies have to be canceled. In particular the mixed anomalies $Z_N SU(3)_c^2$,  $Z_N SU(2)_L^2$ and $Z_N U(1)_Y^2$ anomalies have to vanish modulo $N$. For three generations of neutrinos, a simple candidate anomaly-free symmetry is $Z_3$. Then the lowest order nonperturbative operator breaking
$U(1)$ is
\be 
e^{- 12 \pi \langle t \rangle } c M_P (\Phi^3 + {\bar \Phi}^3) \ = 
e^{- 12 \pi \langle t \rangle } c M_P (2 S^3 - 6 S A^2) \ . \label{zn5} 
\ee 
This will generate a nonperturbative mass for the field $A$
\be 
M_A^2 = 12 ~ c~   v_S M_P  e^{- 12 \pi \langle t \rangle } \ . \label{zn6}
\ee
For moderate values of $\langle t \rangle$ this generates a large hierarchy for 
\be 
\frac{M_A}{M_S} \sim e^{- 6 \pi \langle t \rangle } \sqrt{\frac{M_P}{v_S}} \ . \label{zn7}
\ee 

\subsection{The Signal}

After the symmetry breaking, the Lagrangian (\ref{Eq:lagrangian4}) becomes
\bea
&&
{\cal L}_\nu= - \frac{h}{\sqrt{2}} s \bar N_2 N_2  + \frac{h \theta}{\sqrt{2}} s
 \left( \bar N_1 N_2 + \bar N_2 N_1  \right)
 \nonumber
 \\
 &&
 -\frac{y_{LR}}{2 \sqrt{2}} H \left( \bar N_1 N_2 + \bar N_2 N_1 \right) - \frac{y_{LR} \theta}{ \sqrt{2}} H\left(\bar N_2 N_2 - \bar N_1 N_1  \right)
 \nonumber
 \\
 &&
 - \frac{m_1}{2} \bar N_1 N_1 - \frac{m_2}{2} \bar N_2 N_2
 \label{Eq:lagrangian6}
 \\
 &&
 -i \frac{h}{\sqrt{2}} A \left( \bar N_2 \gamma^5  N_2 - \theta \bar N_1 \gamma^5 N_2 - \theta \bar N_2  \gamma^5 N_1 + \theta^2 \bar N_1 \gamma^5 N_1  \right) .
 \nonumber
\eea
For $M_A \lesssim 8$ TeV, the dominant decay mode is the 2-body process $A \rightarrow N_1 N_1$ and the width of the $A$ 
boson is now given by
\be
\Gamma_A = \frac{10^{-38}}{4 \pi}  \left( \frac{m_1}{\mrm{1~eV}} \right)^2 \left( \frac{10^{10}~\mrm{GeV}}{v_S}  \right)^2
M_A~
\nonumber
\ee
implying
\be
\tau_A \sim 8 \times 10^{14}  \left( \frac{1 ~\mrm{eV}}{m_1} \right)^2 \left( \frac{v_S}{10^{10} ~\mrm{GeV}} \right)^2
\left( \frac{1~\mrm{GeV}}{M_A} \right)~~\mrm{[s]} ,
\label{Eq:taur212}
\ee
Since $M^R$ is now proportional to $h v_S$, the decay rate becomes independent of $h$ if
we express it in terms of $v_S$ and we are forced to consider sub-PeV masses for
our dark matter candidate.
 As one can see, for relatively light Goldstone masses of order 1 GeV, one can 
 obtain lifetimes in excess of the age of the Universe for $v_S > 10^{11}$ GeV,
making this Goldstone mode an interesting dark matter candidate. 

\begin{figure}
    \begin{center}
    \includegraphics[width=3.in]{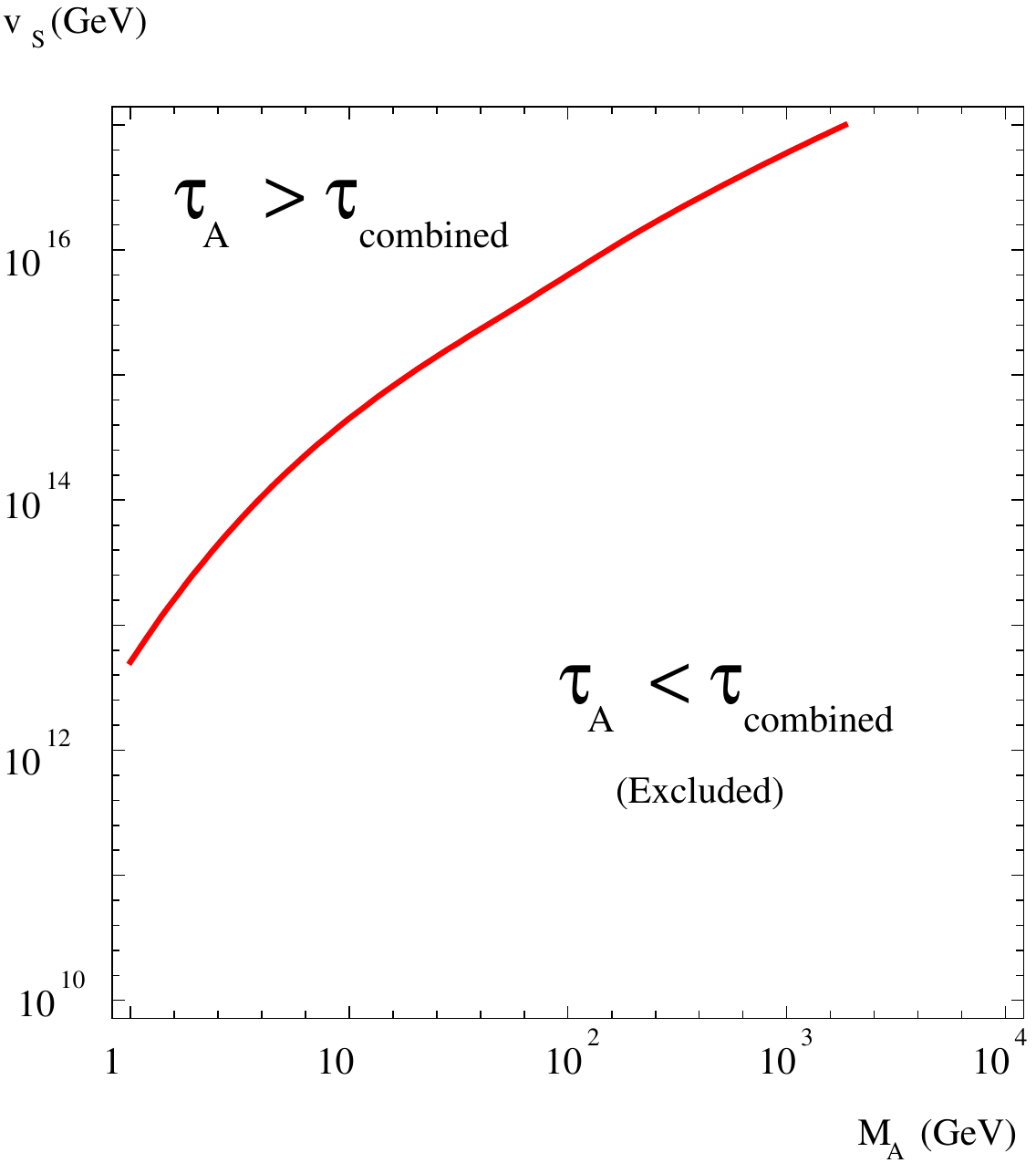}
              \caption{{\footnotesize
       {Parameter space allowed by SuperK and Icecube, in the plane ($M_A$, $v_S$)  for $m_1=1$ eV.
         }
         }}
\label{Fig:vs}
\end{center}
\end{figure}

We show in Fig. \ref{Fig:vs}  the parameter space
allowed by the (non)--observation of signals in neutrino telescope. 
As surprising as it seems, we obtain quite reasonable
values for $v_S$, compatible with GUT--like constructions.

\section*{Conclusion}

In this work, we have shown that a dynamical model to generate majorana neutrino masses, naturally leads to the presence
of a heavy quasi--stable pseudo--scalar particle that can fill the dark matter component of the Universe and whose main decay
mode into two ultra energetic neutrino is a clear signature observable by the IceCube detector. Our work is very general and can be embedded
in many grand-unified scenarios where the breaking of hidden symmetries appears at an intermediate scale.

\noindent {\bf Acknowledgements. }  The authors would like  to thank  S. Pukhov
 and E. Bragina for very useful discussions.
This  work was supported by the Spanish MICINN's
Consolider-Ingenio 2010 Programme  under grant  Multi-Dark {\bf CSD2009-00064}, and the contract {\bf FPA2010-17747}
and the France-US PICS no. 06482.
E.D. and Y.M. are grateful to the Mainz Institute for Theoretical Physics
(MITP) for its hospitality and its partial support during the completion of this
work. Y.M.  acknowledges partial support from the European Union FP7 ITN INVISIBLES (Marie
Curie Actions, PITN- GA-2011- 289442) and  the ERC advanced grants  
 Higgs@LHC. E.D. acknowledges  the ERC advanced grants MassTeV. The work of K.A.O. was supported in part
by DOE grant DE--SC0011842 at the University of Minnesota.

\end{document}